\documentclass[12pt]{article}
\usepackage{latexsym,amsfonts,amsthm,amsmath,amscd,amssymb,color,mathtools}
\usepackage{graphicx}
\usepackage{tikz}
\usetikzlibrary{automata,arrows,positioning,calc}
\usepackage{hyperref}
\usepackage[utf8]{inputenc}
\usepackage{comment}
\usepackage{a4wide}
\usepackage{caption}
\usepackage{float}
\usepackage{pdfpages}

\newtheorem{definition}{Definition}
\newtheorem{theorem}{Theorem}[definition]

\begin{document}
\thispagestyle{empty}
\vspace*{-1.5cm}

\begin{flushright}
  {}
\end{flushright}

\vspace{1.5cm}

%%%%%%%%%%%%%%%%%%%%%%%%%%%%%%%%%%%%%%%%%%%%%%%
%%%%%%%%%%%%%%%%%%%%%%%%%%%%%%%%%%%%%%%%%%%%%%%

\begin{center}
{\huge Unipolar and bipolar aerosol charging as time continuous Markov processes}

\end{center}

%%%%%%%%%%%%%%%%%%%%%%%%%%%%%%%%%%%%%%%%%%%%%%%
%%%%%%%%%%%%%%%%%%%%%%%%%%%%%%%%%%%%%%%%%%%%%%%

\vspace{0.4cm}

\begin{center}
{\bf   Andreas Deser, Jens Kuhne}
 
\end{center}

%%%%%%%%%%%%%%%%%%%%%%%%%%%%%%%%%%%%%%%%%%%%%%%
%%%%%%%%%%%%%%%%%%%%%%%%%%%%%%%%%%%%%%%%%%%%%%%

\vspace{0.4cm}

\begin{center}
  \small{German Federal Office for Radiation Protection \\ 
    Department ``Effects and risks of ionizing and nonionizing radiation''\\
    Ingolstädter Landstraße 1\\
    85764 Oberschleissheim\\
    Germany\\}

  \vspace{0.4cm}

\small{Email: {\tt adeser@bfs.de} and {\tt jkuhne@bfs.de} }

\vspace{0.25cm}

\end{center} 

\vspace{0.4cm}

%%%%%%%%%%%%%%%%%%%%%%%%%%%%%%%%%%%%%%%%%%%%%%%
%%%%%%%%%%%%%%%%%%%%%%%%%%%%%%%%%%%%%%%%%%%%%%%
\begin{center}
\today
\end{center}

\begin{abstract}
 The acquisition of charge by aerosol particles is well-known to be stochastic in nature. We review the principles of charging using the conceptually and computationally clear language of continuous time Markov processes. A novel numeric approach is presented that can be used to calculate the time evolution of various particle charging processes. Its modular character makes it easy to implement and allows for quick adaptation to specific problems. We conclude with the application of ergodicity for finite state-space Markov processes in order to determine stationary charge distributions in case of bipolar charging. 
\end{abstract}

\clearpage

\tableofcontents

\section{Introduction}

%High voltage power transmission lines are sources of ionized air, also referred to as corona ions. They potentially charge aerosols naturally existing in the surrounding air. Accumulation of charge due to inhalation of aerosols into the lung is a potential health risk, the careful estimation of the amount of charged aerosols in the vicinity of high voltage power transmission lines is therefore of special interest.

The stochastic nature of acquisition of electrical charge by aerosol particles is well known \cite{boisdron1970stochastic, marlow1975calculations}. In the simplest model, two distinct steps determine the rate of aerosol charging: First, the purely stochastic process of a charged particle meeting an aerosol of a specific initial charge. And second the charging process itself. The latter is governed by diffusion of electrically charged particles through a spherical shell around the aerosol particles in the background of the electrical field produced by the initial charge of the aerosol. The rate of charging can be modeled by acquisition coefficients depending on $k_B T$ as well as on the size and charge of the aerosol \cite{gunn1954diffusion, fuchs1963stationary, wiedensohler1988approximation, wiedensohler1990bipolar}.

In this work, we focus on the stochastic process determining the first step mentioned above. Whereas in \cite{boisdron1970stochastic}, integration of a Boltzmann-type equation resulted in ordinary differential equations for the change of concentration of aerosols of different charge, here we first identify the stochastic process of charging as a specific time continuous Markov process. The idea is to regard aerosol charging as a Poisson process whose time constant changes whenever charge is captured. We also allow for an arbitrary initial distribution of charges among the aerosols as well as acquiring different amounts of charge in one step.  Using this approach, the equations given in \cite{boisdron1970stochastic} are a direct consequence of the Markov process: The Kolmogorov forward equations. The purpose of our approach is logical transparency which allows a conceptually clear way to generalize to more complicated situations such as time-varying acquisition coefficients. Furthermore we develop a numerical MatLab routine to implement different situations in a convenient way. Finally, the Markov process formalism allows for universal statements in aerosol charging, such as the existence and uniqueness of steady-state distributions. These follow directly by applying existing ergodic theorems for Markov processes.

The paper is organized as follows. In section \ref{review} we give a brief review of basic notions of Markov process theory. This is textbook knowledge and we provide our personal choice of literature. In section \ref{UnipolarMarkov} we identify the type of Markov process relevant for aerosol charging. In order to keep the notation simple, we discuss the bipolar case in detail and mention the unipolar and general multipolar cases as useful applications. Finally the concept of ergodicity is applied to the bipolar and multipolar cases to show the existence of steady-state charge distributions. In section \ref{Numerik} we apply a numerical study of the derived concept to show the advantage of the formalism in order to establish realistic modeling of aerosol charging. An appendix containing a MatLab code is provided.  

\section{A brief review of time continuous Markov processes}
\label{review}
Time continuous Markov processes are among the most studied stochastic processes and are widely used in many fields as diverse as probability theory, statistical mechanics and economy. They describe parameterized families of events (random variables) whose future only depends on their current state. This is called Markov property. In order to set up the basic notation, we briefly review elementary aspects of Markov processes needed to formalize aerosol charging. For a systematic introduction to the theory we refer to the literature on the subject, we used in particular the textbooks \cite{georgii2015stochastik, klenke2006wahrscheinlichkeitstheorie} and references therein.
\begin{definition}
  Let $X_t$ be a family of random variables, parameterized by $t\in [0,\infty)$ and taking values in a countable set $E$. $X_t$ is called \emph{time continuous Markov process}, if
    \begin{align}
      P(X_{s+t} = j | X_{s_1}=k_1, X_{s_2} = k_2,\dots, X_{s_n} = k_n, X_s = i) =&\, P(X_{s+t}=j | X_s = i) \nonumber \\
      =&\, \Pi_t(i,j)\;,
    \end{align}
    for all $t>0,\, s > s_n > \dots > s_1 \geq 0$, with $i,j,k_1,\dots,k_n \in E$ and a stochastic matrix $\Pi_t$. 
\end{definition}
The countable set $E$ is usually termed state space. Only the conditional probability $P(X_{s+t}=j | X_s = i)$ is relevant, and we assume that it only depends on the time difference $t$ (time homogeneity). The quadratic matrix $\Pi_t$ is stochastic, i.e. its row elements are either zero or positive and add up to $1$. Its dimension is determined by the number of elements of the state space $E$. It is straightforward to see that one has the semi-group property of time evolution $\Pi_{s+t} = \Pi_t \Pi_s$ and initially $\Pi_0 = \mathbf{1}$, where $\mathbf{1}$ is the identity matrix. Important information about the process is obtained by looking at its evolution in infinitesimal time:
\begin{definition}
  Let $X_t$ be a continuous Markov process. The \emph{infinitesimal generator of the process} is defined by 
     \begin{equation}
    G(i,j) := \, \underset{h \stackrel{>}{\rightarrow} 0}{\lim} \, \frac{\Pi_h(i,j) - \mathbf{1}(i,j)}{h}\;.
  \end{equation}
\end{definition}
\noindent By writing $\mathbf{1}(i,j)$ we assume the rows and columns of the identity to be indexed by the elements of $E$ and $\mathbf{1}(i,j)$ is the matrix entry at position $(i,j) \in E \times E$. The following properties of the infinitesimal generator are of importance for our work. Their proofs are straightforward and can be found in textbooks on the subject.
\begin{itemize}
\item $\sum_j G(i,j) = 0$, i.e. the sums over row elements of $G$ vanish.
\item The Kolmogorov equations hold, $\tfrac{d}{dt} \Pi_t = G \Pi_t$ and $\tfrac{d}{dt} \Pi_t = \Pi_t G$.
\item The Kolmogorov equations are solved by $\Pi_t(i,j) = e^{tG}(i,j)$, where as usual the matrix exponential is defined by the power series $e^{tG}(i,j) = \sum_{n = 0}^\infty\,\frac{t^n}{n!}\,G^n(i,j)$\;.
\end{itemize}
Once the state space and generator are identified, various probability distributions of the process are calculated using the solution to the Kolmogorov equations. Let $X_t$ be the random variable with values in $E$, the probability of $X_t = j$ when starting at $X_0 = i$ is given by\footnote{We use the popular notation $P^i(X_t = j)$ for the probability of $X_t = j$, when starting at $i \in E$, i.e. $P(X_t = j | X_0 = i)$.}
\begin{equation}\label{CentralProbability}
  P^i(X_t = j) = \; e^{t G}(i,j)\;,\quad \textrm{for} \; i,j \in E\;.
\end{equation}
It is easily seen that, given an initial distribution $\mathbf{x}_0 = (x_{k_0},x_{k_1},\dots)$, for $k_l \in E$ and $\underset{k_l \in E}{\sum}\,x_{k_l} = 1$, the distribution at time $t>0$ is calculated by applying the exponential of the generator:
\begin{equation}\label{CPcons}
  \mathbf{x}(t) = \, \mathbf{x}_0\,\Pi_t =\, \mathbf{x}_0\,e^{tG}\;,\quad \mathbf{x}(t) = (x_{k_0}(t), x_{k_1}(t), \dots)\;.
\end{equation}
Using this, the Kolmogorov equations translate in a set of first order linear ordinary differential equations:
\begin{equation}\label{ode}
  \frac{d}{dt} \,\mathbf{x}(t) = \mathbf{x}_0 \,e^{tG} \,G = \mathbf{x}(t)\,G\;,
\end{equation}
so in short $\tfrac{d}{dt} \mathbf{x} = \mathbf{x}\,G \;, \,\mathbf{x}(0) = \mathbf{x}_0$. 
Typical examples of Markov processes appear in queuing theory, birth-death processes or population dynamics. In case of discrete time, random walks on lattices are characteristic examples. A very basic example of a continuous time Markov process is the Poisson process, which we now define and review in the terminology of Markov processes. This is the starting point for generalizations and applications to aerosol charging in the next section. Let $(L_i)_{i\geq 0}$ be a series of independent random variables, exponentially distributed with common parameter $\mu \in \mathbb{R}$, i.e. with distribution function $\rho_{L_i}(t) = \mu\,e^{-\mu t}$. Furthermore, let $T_k := \sum_{i=1}^k \, L_i$ be the time of the $k$'th event happening. One can then show that the number of events $X_t$ that happened until some fixed time \footnote{More precisely, $X_t := \underset{k\geq 1}{\sum}\, 1_{\left]0,t\right]}(T_k)$, with $1_{A}$ the characteristic function of a subset $A\subset \mathbb{R}$.} $t$ is Poisson distributed with parameter $\mu t$, i.e.
    \begin{equation}\label{elemPoisson}
      P(N_t = k) = \, \frac{(\mu t)^k}{k!}\,e^{-\mu t}\;.
    \end{equation}
    As for this process the infinitesimal rate of change from the state $N_t = k$ to $N_t = k+1$ is $\mu$, it is clear how to construct the generator of the corresponding Markov process. As we count events starting from $0$, the state space is $E = \mathbb{Z}_0^+$ and 
    \begin{equation}\label{Poissongenerator}
    G(i,j) = \begin{pmatrix}
      -\mu & \mu & 0 & 0 &\dots\\
      0 & -\mu & \mu & 0 &\dots\\
      0 & 0 & -\mu & \mu &\dots\\
      \dots & & & &
    \end{pmatrix}\;, \quad i,j \in E\;.
  \end{equation}
    Indeed, using equation \eqref{CentralProbability} for the Markov process the probability of $k$ events happening, starting from $k=0$ is
    \begin{equation}
      P^0(X_t = k) =\, \left( e^{tG} \right)_{0,k} =\, \frac{(\mu t)^k}{k!}\,e^{-\mu t}\;,
    \end{equation}
    coinciding with \eqref{elemPoisson}. The probability of $j$ events happening, starting at event $i$ is computed analogously for $i,j \in E$. As the rate $\mu$ is not changing in the whole process, this model clearly is too easy to describe aerosol charging, but conceptually it is now immediate how to generalize.

    Finally we state a basic result of ergodic theory for continuous Markov chains. Intuitively, if for sufficiently large times every state can be reached from any other state in a \emph{finite} state space, there is hope to reach a steady state distribution. The mathematical condition for the latter is irreducibility of the generator. $G$ is called irreducible, if for all $i,j \in E$ one can find an $N \in \mathbb{N}$ and $k_0, k_1, \dots k_N \in E$ with $k_0 = i$ and $k_N = j$ such that $\prod_{l=1}^N\, G(k_{l-1},k_l) \neq 0$. For irreducible generators one has the following basic result:
    \begin{theorem}\label{ergodenthm}
      Let $E$ be a finite state space and $G$ an irreducible generator of a continuous Markov process and $\Pi_t(i,j) = e^{tG}(i,j)$. Then $\lim_{t\rightarrow \infty}\, \Pi_t(i,j) = \boldsymbol\alpha(j)$ independent of $i\in E$. Furthermore, $\boldsymbol\alpha$ is the only probablility distribution on $E$ which satisfies $\boldsymbol\alpha \,G = 0$, or equivalently $\boldsymbol\alpha\,\Pi_s = \boldsymbol\alpha$ for all $s>0$.
    \end{theorem}
    For the proof we refer to the textbooks on the subject, e.g. \cite{georgii2015stochastik, klenke2006wahrscheinlichkeitstheorie}. Realistically, any charging process for aerosol particles has a finite state space, as it is not possible to infinitely charge an aerosol particle. In case of an irreducible generator theorem \ref{ergodenthm} then guarantees the existence of a steady state distribution and gives a simple recipe to calculate it. 
    
    \section{Unipolar and bipolar aerosol charging as Markov processes}
    \label{UnipolarMarkov}
    Using the notation introduced in the previous section, we identify aerosol charging as continuous Markov processes. We start with the general case of bipolar charging with different concentrations of positive and negative ions and treat unipolar charging as a special case. After commenting on various possibilities for generalizations we apply ergodicity to prove the existence of a steady state charge distribution for the bipolar case. 

    \subsection{Bipolar charging with different ion concentrations}
    We begin with the idealized situation that arbitrary positive and negative charging of aerosol particles by the acquisition of positively as well as negatively charged ions is possible. To simplify the identification of the Markov process, we assume that there are only ions with one positive or one negative elementary charge. The concentration as well as mobility of the two types of ions can be different. Finally we assume that the aerosol particle can not loose ions, therefore its charge increases by acquiring a positive ion and decreases by the acquisition of a negative ion. Hence, in case of arbitrary possible charging of an aerosol, the state space is given by integer numbers $E = \mathbb{Z}$. In contrast to the Poisson process, the rate of acquisition of an ion depends on the current charge state. Furthermore there are different rate constants $(\mu_i)_{i\in \mathbb{Z}}$ and $(\eta_i)_{i\in \mathbb{Z}}$ for positive and negative ions, respectively. By adjusting the generator \eqref{Poissongenerator} for the Poisson process accordingly we get for the generator of bipolar charging:
    \begin{equation}
      \label{gbi}
    G_{bi}(i,j) = \begin{pmatrix}
      \dots & & & & & & & \\
      \dots & \eta_{-1} & -(\eta_{-1} + \mu_{-1}) & \mu_{-1} & 0 & 0 & 0 & \dots\\
      \dots & 0 & \eta_0 & -(\eta_0 + \mu_0) & \mu_0 & 0 & 0 & \dots\\
      \dots & 0 & 0  & \eta_1 & -(\eta_1 + \mu_1) & \mu_1 & 0 & \dots\\
      & & & & & & & \dots
    \end{pmatrix}\;.
    \end{equation}
    The rows and columns of the generator extend over the integer numbers. The rate constants for acquiring a positively charged ions appear in the upper off-diagonal and the rate constants for acquiring a negatively charged ion in the lower off-diagonal, respectively. To implement the fact that positive and negative ions come in different concentrations, we take advantage of the fact that $G_{bi}$ can be written as the sum of two matrices $G_{bi} = \mathbf{\eta} + \mathbf{\mu} $, for
      \begin{equation}
        \mathbf{\eta}= \begin{pmatrix}
          \dots & & & & & & \\
          \dots & \eta_{-1} & -\eta_1 & 0 & 0 & 0 & \dots \\
          \dots & 0 & \eta_0 & -\eta_0 & 0 & 0 &\dots \\
          \dots & 0 &  0 & \eta_1 & -\eta_1 & 0 & \dots \\
          & & & & & & \dots
        \end{pmatrix}\;,\;\mathbf{\mu} = \begin{pmatrix}
          \dots & & & & & & \\
          \dots & 0 & -\mu_{-1} & \mu_{-1} & 0 & 0 & \dots \\
          \dots & 0 & 0 & -\mu_0 & \mu_0 & 0 & \dots \\
          \dots & 0 & 0 & 0 & -\mu_1 & \mu_1 & \dots \\
          & & & & & &\dots
        \end{pmatrix}
      \end{equation}
      Implementing different initial concentrations $n_-$ and $n_+$ for positive and negative ions, respectively, is done by multiplying the corresponding matrix, i.e. $G_{bi} = n_-\, \mathbf{\eta} + n_+\,\mathbf{\mu}$. 

      In practice, the existence of arbitrary high positively or negatively charged aerosol particles is highly unlikely. It is more realistic to assume a maximally positive charge and a minimally negative charge a particle can acquire. It turns out that typically charge numbers from $-20$ to $20$ appear and the probabilities to charge more positive or more negative can be assumed to be zero. This means that the state spaces will be finite (e.g. $E = \{-20, -19, \dots, 19, 20\}$, the matrix $\mu$ have a zero column for maximally positive charge and the matrix $\eta$ will have a zero row for maximally negative charge. As an example, in case fo $E = \{-2,-1,0,1,2\}$ and $n_+ = n_1 = 1$, without loss of any generality the generator for bipolar charging is
      \begin{equation}\label{bipolargenerator}
        G_{bi} = \begin{pmatrix}
          -\mu_{-2} & \mu_{-2} & 0 & 0 & 0 \\
          \eta_{-1} & -(\eta_{-1} + \mu_{-1}) & \mu_1 & 0 & 0 \\
          0 & \eta_0 & -(\eta_0 + \mu_0) & \mu_0 & 0 \\
          0 & 0 & \eta_{1} & -(\eta_1 + \mu_1) & \mu_1 \\
          0 & 0 & 0 & \eta_2 & -\eta_2
        \end{pmatrix}\;.
      \end{equation}
      It is clear how the structure looks for general finite state spaces $E = (-n, -n+1, \dots, n-1, n)$. It is not necessary to have the states symmetrically distributed around $0\in E$, but for convenience we will focus on this case. For completeness, the Kolmogorov equations in the form \eqref{ode} for a given initial probability distribution $\mathbf{x}_0 = (x_{-n},x_{-n+1},\dots,x_{n-1},x_n)$ are given below, using Newton's notation for time derivatives:
      \begin{align}\label{kolmogorov}
        \dot{x}_{-n}(t) =\,& -\mu_{-n} x_{-n}(t) + \eta_{-n+1}\,x_{-n+1}(t)\;,\nonumber \\
        \dot{x}_{-n+1}(t) =\,& \mu_{-n}\,x_{-n}(t) -(\eta_{-n+1} + \mu_{-n+1})x_{-n+1}(t) + \eta_{-n+2}\, x_{-n+2}(t) \;,\nonumber \\
        \dot{x}_{-n+2}(t) =\,& \mu_{-n+1}\,x_{-n+1}(t) -(\eta_{-n+2} +\mu_{-n+2}) x_{-n+2}(t) + \eta_{-n+3}\,x_{-n+3}(t)\;,\nonumber \\
        &\dots \nonumber \\
        \dot{x}_{n-1}(t) =\,& \mu_{n-2}\,x_{n-2}(t) - (\eta_{n-1}+\mu_{n-1}) x_{n-1}(t) + \eta_n\,x_n(t)\nonumber \\
        \dot{x}_n(t) =\,&\mu_{n-1}\,x_{n-1}(t) - \eta_n\, x_n(t)\;.
      \end{align}
      Systems of ODE's of this type were studied earlier in the literature for various scenarios of particle charging \cite{boisdron1970stochastic}. The Kolmogorov equations thus establish the link between the Markov process formalism to earlier work on the subject. 
        
      We will use  matrices of the form \eqref{bipolargenerator} combined with concentrations $n_-, n_+$ for detailed numerical studies in section \ref{Numerik}. Before doing so we will look at variations of the formalism to treat other situations and investigate the steady state behaviour of bipolar charging. 
    
      \subsection{Other scenarios: Unipolar charging and charging with multiple valued ions}
      An advantage of the formalism used so far is its conceptual simplicity. In particular it allows for an easy implementation of different situations frequently used in aerosol physics. A case of interest is \emph{unipolar} charging, where there are only ions of one type. It is easy to see that setting either $n_- = 0$ or $n_+ = 0$ gives the desired results, as this sets either the contribution of $\eta$ or $\mu$ to the generator to zero. In case of a state space $E = \{-2, -1, 0, 1, 2\}$ this means
      \begin{equation}
        G_{uni} = \begin{pmatrix}
          -\mu_{-2} & \mu_{-2} & 0 & 0 & 0 \\
          0 & -\mu_{-1} & \mu_{-1} & 0 & 0\\
          0 & 0 & -\mu_0 & \mu_0 & 0 \\
          0 & 0 & 0 & -\mu_1 & \mu_1 \\
          0 & 0 & 0 & 0 & 0
        \end{pmatrix}\;.
      \end{equation}
      We note that the zero row in $G_{uni}$ will result in a row $(0,0,0,0,1)$ in the corresponding stochastic matrix $\Pi_t = e^{tG}$. This is what is expected: In case of charging without loosing ions and a maximal positive charge value, the probability of staying at the element $-2 \in E$ is $1$. An element of $E$ of this kind is sometimes also called \emph{absorbing state}.

      The formalism in addition allows for the extension to charging by multiply valued ions. To illustrate this without overloading the notation, we assume again a state space of $E = \{-2, -1, 0, 1, 2 \}$ and the existence of ions which carry one or two positive charges and one or two negative charges. The respective charging rates are dentoted by $\mu_i ^+, \mu_i^{++}$ as well as $\eta_i^-, \eta_i ^{--}$, respectively. In case of all concentrations equal, we therefore get for the generator
      \begin{equation}\label{alles}
        G_{mult} =
         \begin{pmatrix}
          -\Sigma_{-2} & \mu_{-2}^+ & \mu_{-2}^{++} & 0 & 0 \\
          \eta_{-1}^- & -\Sigma_{-1} & \mu_{-1}^+ & \mu_{-1}^{++} & 0 \\
          \eta_0^{--} & \eta_0^- & -\Sigma_0 & \mu_0^+ & \mu_0^{++}\\
          0 & \eta_1^{--} & \eta_1^- & -\Sigma_1 & \mu_1^+\\
          0 & 0 & \eta_2^{--} & \eta_2^- & -\Sigma_2
        \end{pmatrix}\;,
      \end{equation}
      where the $\Sigma_i$ are chosen in such a way to satisfy the properties of a generator, i.e.
      \begin{align}
        \Sigma_{-2} = &\mu_{-2}^+ + \mu_{-2}^{++}\,, \quad \Sigma_{-1} = \eta_{-1}^- + \mu_{-1}^+ + \mu_{-1}^{++}\, \nonumber \\
        \Sigma_{2} = &\eta_2^{--}+\eta_2^{-}\,, \quad  \Sigma_1 = \eta_1^{--} + \eta_1^- + \mu_1^+ \;,\nonumber \\
        &\Sigma_0 = \eta_0^{--} + \eta_0^- + \mu_0^+ + \mu_0^{++}\;.
      \end{align}
      We observe that e.g. $\eta_{-1}^{--}$ does not exist due to our assumptions on the state space. Similarly for $\mu_1^{++}$ and so forth. Furthermore, the implementation of different concentrations of the differntly charged ions is done analogously to the bipolar case: The matrix \eqref{alles} can be decomposed in matrices containing the rates for the different types of ions: $G_{mult} = \eta^{--} + \eta^- + \mu^+ +\mu^{++}$. Multiplying each matrix with the corresponding concentration gives the generator for the entire process:
      \begin{equation}
        G^{full}_{mult} = n_{--}\,\eta^{--} + n_- \,\eta^- + n_+ \,\mu^+ + n_{++}\,\mu^{++}\;.
      \end{equation}
As a result, our formalism easily allows to treat aerosol charging with ions carrying different charge values and existing in differnt concentrations and with or without a maximal or minimal total charge of the aerosol particle. 
        
    \subsection{Ergodicity for the bipolar and multipolar cases}
    The ergodic theorem for finite state space continuous Markov processes reviewed at the end of section \ref{review} allows to investigate whether there are steady state charge distributions in the various scenarios studied so far. Clearly, for unipolar charging, the generator is not irreducible, as one can not move backwards (i.e. have charge loss) in the charging process. More precisely, starting with a charge $k$ and ending in a charge $l$, for $k>l$, in the product of matrix elements needed for irreducibility, at least one matrix element has to be taken from below the diagonal, but all these elements vanish and hence the product vanishes. However, the generator in the bipolar and multipolar cases has the irreducibility property. As an example, in the bipolar case, using the notation of theorem \ref{ergodenthm} and section \ref{review}, first take $k_0 = k$ and $k_N = l$ for $k,l \in \mathbb{Z}$ and $k<l$. Then for $N \in \mathbb{N}$ such that $k_0 = k, k_1 = k+1, \dots, k_N = l$ one has
    \begin{align}
      \prod_{n=1}^N\, G(k_{n-1}, k_n) =&\, G(k,k+1)G(k+1,k+2)\cdots G(l-1,l)\nonumber \\
      =&\, \mu_k \mu_{k+1} \cdots \mu_l\;.
    \end{align}
    Second, in case $k>l$, taking $N$ such that $k_0 = k, k_1 = k-1, \dots k_N = l$ one has
    \begin{align}
      \prod_{n=1}^N\, G(k_{n-1},k_n) =& \,G(k,k-1)G(k-1,k-2)\cdots G(l+1,l)\nonumber \\
      =&\,\eta_k \eta_{k-1}\cdots \eta_l\;.
    \end{align}
    As we assumed all $\mu_k$ and $\eta_k$ nonvanishing (i.e. nonvanishing charging rates), the products in the last two expressions are nonzero, for all $k,l \in \mathbb{Z}$. Thus, $G$ is irreducible and ergodicity is guaranteed. Note that this is even more evident in the multipolar case. In this case even if some of the $\eta$'s or $\mu$'s may vanish, in case another path using other (multiple charge) rates exists, there still exist stationary charge distributions. All of them can be computed rigorously by solving the corresponding eigenvalue problem mentioned in theorem \ref{ergodenthm}. We will present a numerical study of the results in the next section.
    
    \section{Numerical results}
    \label{Numerik}
    To investigate the applicability of the formalism outlined above, we modeled the charging of particles with a diameter $d$ of $300$nm at moderate temperature ($T=300$K) in a bipolar ion cloud with constant properties using the Markov chain approach developed above. To apply the equations listed in chapter \ref{UnipolarMarkov}, the rate coefficients $\mu_i$ and $\eta_i$ for the uptake of an additional positive or negative elementary charge at any relevant charge number $q_i$ have to be known. Without any loss of generality, the simple analytic formulae derived by Gunn \cite{gunn1954diffusion} are used to calculate $\mu_i$ \eqref{mui} and $\eta_i$ \eqref{etai} although more precise derivations have been published in the past (\cite{fuchs1963stationary, hoppel1986ion}).
    \begin{align}
      \mu_i =&\,\frac{q_i\,e\,m_+}{\epsilon_0(e^{2\lambda_i q_i}-1)}\;, \label{mui}\\
      \eta_i =&\, \frac{q_i\,e\,m_-}{\epsilon_0(1-e^{-2\lambda_i q_i})} \label{etai}\;,
    \end{align}
    with $\lambda = \,\frac{e^2}{4\pi\epsilon_0\,d\,k_B\,T}$.

    Here $e$ represents the elementary charge, $m_{\pm}$ the ion mobility of the positively or negatively charged ion, respectively, $\epsilon_0$ the dielectric permittivity, $d$ the particle diameter ($300$nm), $k_B$ the Boltzmann constant and $T$ the absolute temperature ($300$K).

\begin{figure}[H]
      \centering
      \includegraphics[width=\textwidth]{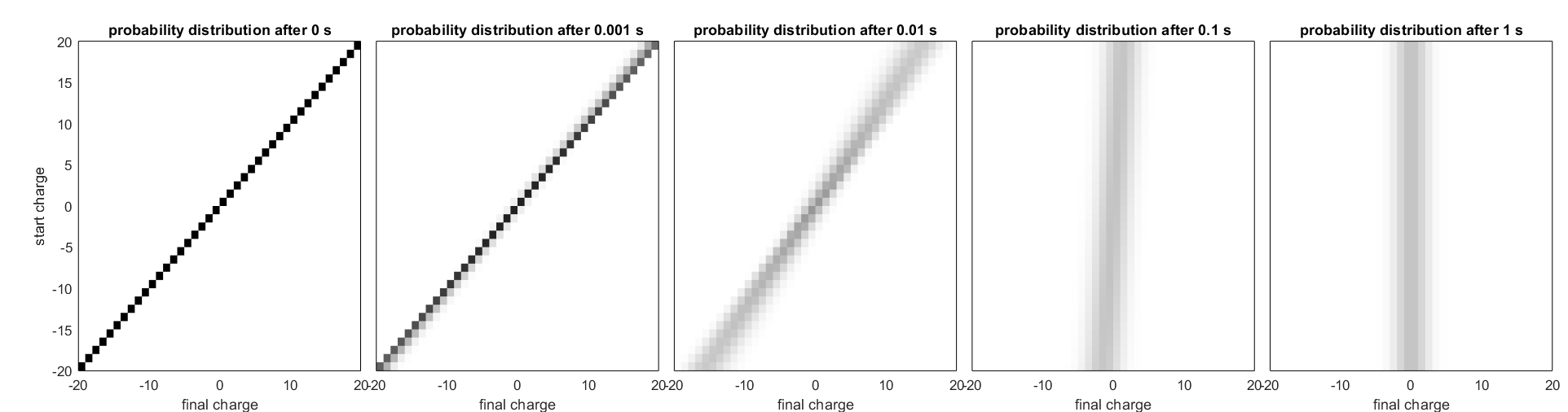}
      \caption{Probability distribution matrices for several time points after charging in a bipolar environment.}
      \label{fig1}
    \end{figure}

    For neutrally charged aerosols, the rate coefficient equations listed above are not valid. For this special case, the rate coefficients are given by
    \begin{equation}
      \mu(q=0) = \,\frac{e\,m_+}{2\epsilon_0 \lambda} \;,\qquad \eta(q=0) =\, \frac{e\,m_-}{2\epsilon_0 \lambda}\;.
    \end{equation}
    In analogy to equations \eqref{gbi}-\eqref{bipolargenerator}, we constructed $G_{bi}(i,j)$ using $\mu_i$ and $\eta_i$ for any integer starting charge numbers ranging from $-20$ to $20$. Assuming a bipolar environment of $10^7$ singly charged ions per $\textrm{cm}^3$ of each polarity as well as ion mobilities of $m_+ =\, 1.35\cdot10^4Vm/s$ and $m_- =\, 1.6\cdot10^{-4}Vm/s$ for positive and negative ions (taken from \cite{WIEDENSOHLER1986413}), respectively, we calculated the probability distribution matrix $\Pi_t = e^{tG}$ using equation \eqref{CentralProbability} for $5$ different time points (see figure \ref{fig1}). Matlab code examples are given in the Annex.

    The $i$-th row of each probability distribution matrix represents the charge probability distribution that would be expected for the final state after charging aerosols with start charge number $q_{si}$ for $t$ seconds. The probability distribution matrix at time point $0$ ($\Pi_{t=0}$) is a diagonal matrix indicating that the charge distribution is not altered after zero time. With increasing time, the probability distribution matrix gradually approaches a state, where the probability distribution is independent from the start charge $q_{si}$ of the aerosol.

\begin{figure}[H]
      \centering
      \includegraphics[width=\textwidth]{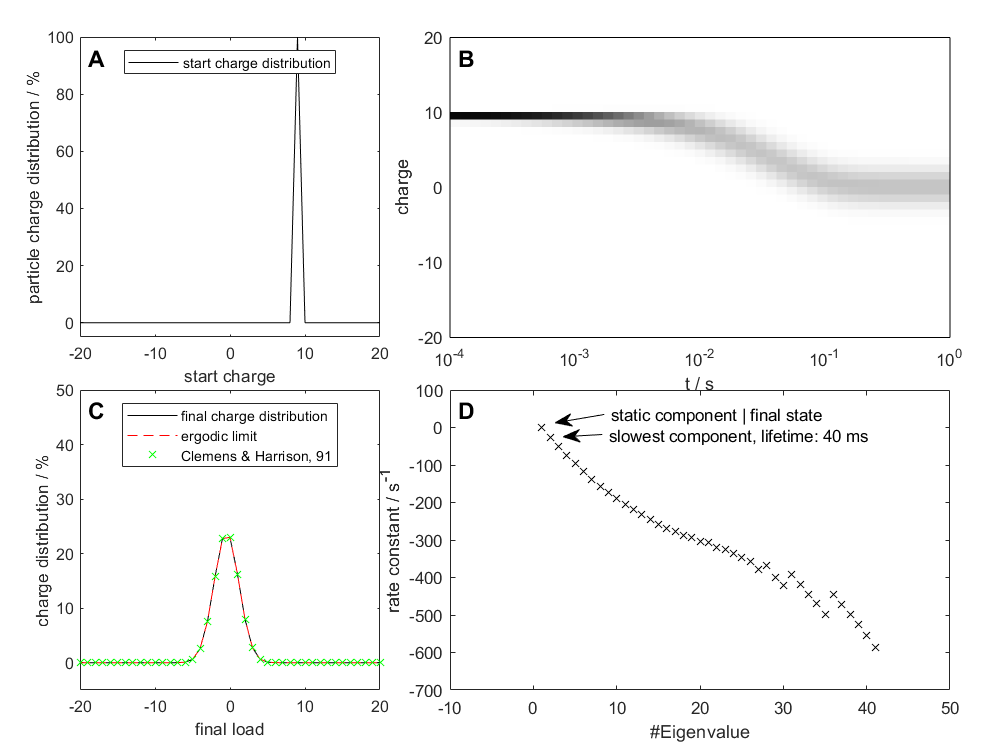}
      \caption{{\bf Time evolution of the charging process.} A, start charge distribution containing a single component loaded with nine elementary charges. B, time evolution of the charging process. C, final charge calculated by different methods: Distribution after $1$s (black), ergodic limit according to theorem \ref{ergodenthm} (red) and the analytic solution (Clemens and Harrison '91). D, rate constants given by the eigenvalues of the transposed generator matrix. The slowest process involved in the charging process exhibits a rate constant of approx. $25/s$ (lifetime $= 40$ ms) }
      \label{fig2}
    \end{figure}

    The charge distribution after charging an aerosol with given initial aerosol charge distribution $x_0$ for $t$ seconds is calculated by applying equation \eqref{CPcons}. Technically, this procedure corresponds to generating a charge distribution matrix by weighting the rows of the probability distribution matrix with the respective components of the initial charge distribution of the aerosol and summing along its columns. Applying this approach for several time points, we calculated the time evolution of the charging process in a bipolar environment for an arbitrary initial charge distribution (Fig. \ref{fig2}).

    \vspace{8pt}
    
    \vspace{8pt}
    
    The arbitrary initial charge distribution shown in Fig. \ref{fig2}A contains a distinct aerosol component loaded with nine positive elementary excess charges. The charging process in the bipolar environment leads to a slow change in charge distribution until an equilibrium steady state charge distribution is reached after approximately $10^{-1}$s (Fig. \ref{fig2}B). The final charge distribution after a charging time of $1$s is shown in Fig. \ref{fig2}C (black line). This distribution is equivalent to the ergodic limit of the steady state charge distribution that is given by the kernel of the matrix $G$ as outlined in theorem \ref{ergodenthm} (see Matlab code example in the Annex for details).

    Both, the final charge distribution after $1$s and the ergodic limit of the steady state charge distribution are consistent with the analytic solution for the steady state charge distribution suggested by \cite{clement1992charging} (Fig. \ref{fig2}C, green X).

    The charging process occurs on several timescales. From the Kolmogorov equations, i.e the system of linear ODEs \eqref{ode}, \eqref{kolmogorov} follows that the time evolution of the charge number population can be expressed by a weighted sum of exponentials whose rate constants are certain combinations of the entries in the generator matrix. More precisely, these rate constants are given by the eigenvalues of the transposed generator matrix (see Fig. \ref{fig2}D) and are therefore independent from the initial charge distribution. The eigenvalue ``$0$'' (corresponding to $0/s$) describes a time independent component, namely the final state (Fig. \ref{fig2}C). The next higher rate constant describes the slowest process involved in charging and is therefore a measure of the velocity of the overall charging process. The corresponding lifetime of this slowest process (given by the negative inverse rate constant) is approximately $40$ ms for the example discussed here. This means that for the conditions of the assumed bipolar environment (i.e. a bipolar neutralisator), a charging time of less than a second would be sufficient to reach the final charge distribution.
    
    %The charging process occurs on several timescales. Because the starting charge distribution is far from equilibrium, the depopulation of the initial state ($+9$) closely resembles a single exponential process with a lifetime in the order of $10^7-10^8$s (Fig. \ref{fig2}D, orange line). The final charge distribution is reached after times $> 10^9$s (Fig. \ref{fig2}D, blue line). Aerosols with excess charges that are higher than the ones of the final distribution but lower than the one of the initial state accumulate and depopulate during the charging process. For instance, particles with $+4$ excess charges reach a maximum concentration after $2\cdot 10^8 - 3\cdot 10^8$s but are not present at the beginning and the end of the charging process (Fig. \ref{fig2}D, red line).

\begin{figure}[H]
      \centering
      \includegraphics[width=0.9\textwidth]{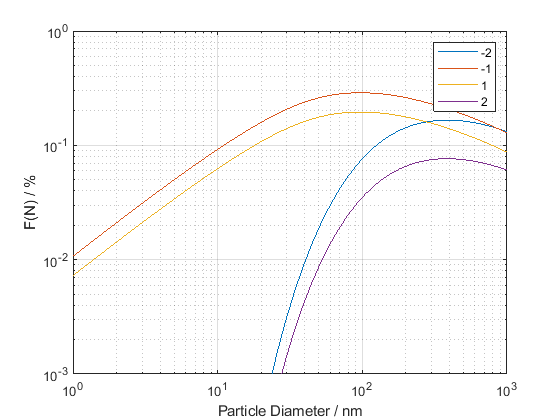}
      \caption{{\bf Bipolar charge distribution for different particle size.} The overall shape of the curves resemble the ones published in \cite{WIEDENSOHLER1986413}, more realistic charge distributions can be derived by replacing the simple attachment coefficients by more precise ones that take additional physical properties into account.}
      \label{fig3}
    \end{figure}

    In analogy to results published earlier using Fuchs theory \cite{WIEDENSOHLER1986413}, size dependent final bipolar charge distributions can be calculated by varying the parameter $d$ in the Markov chain approach (Fig. \ref{fig3}). By using more precise attachment coefficients than done in our study, comparing theoretical values with experiment becomes feasible as well.
    
    \section{Conclusion and outlook}
    Using time continuous Markov processes, we studied unipolar, bipolar and multipolar aerosol charging processes. This conceptually simple and numerically powerful viewpoint potentially allows for calculating charging probability distributions known in the literature by using experimentally determined charging rates. One of the advantages of the formalism is its structural clarity and the use of powerful general results such as the ergodicity theorem for finite state space Markov chains. The latter e.g. allows for a numerically fast computation of limiting steady state charge distributions.

    Another advantage of the formalism is its extendability to more complicated (realistic) charging processes. Aerosol clouds with particles of multiple diameters and varying initial charge distributions can be treated as well as arbitrary mixture of multiply charged ion types, each coming with its own concentration. Further generalizations not treated in this work are time-varying concentrations usually found in situations where the number of ions continuously decreases when there is no infinite supply \cite{hoppel1986ion} or ions are created by an experimental setup. In our treatment it is clear how to implement such cases. The mathematical theory behind these processes is more complicated, as the stochastic matrix used in the Markov property will become time-dependent in the sense that not only differences play a role, but also the time of the event one is looking at. The theory of time-inhomogenious continuous Markov processes provides the correct setup to treat these cases, including more intricate ergodicity statements.

\subsection*{Acknowledgements}
The authors are indepted to Ulf Winkler and Alfred Wiedensohler for inspiring discussions on the ideas of aerosol charging.

\bibliography{ref}
\bibliographystyle{utphys}



\section*{Supplementary material}

# Annex: Matlab examples for numeric implementation of the Markov chain approach to model unipolar and bipolar charging

1. Matlab function to calculate the generator and plot the bipolar charging probability matrix (ProbMatrix) for given ion concentration of positive (n_pos) and negative ion concentration (n_neg), arrays of attachment coefficients for positive (mu) and negative (nu) ions, charge number range (k) and time point (t):

```matlab
function [ProbMatrix,Generator]=GeneratorConstructionFunction(n_pos,n_neg,mu,nu,k,t)
%% Construction of Generator Matrix G using diagonal matrices of ion concentration weighted
%% attachment coefficients
n=length(k);
mudiag=n_pos*diag(mu);
nudiag=n_neg*diag(nu);
Generator=-(mudiag+nudiag)+[zeros(n,1),mudiag(:,1:(n-1))]+[nudiag(:,2:(n)),zeros(n,1)];
%% satisfying generator properties at boundaries
Generator(1,1)=-mudiag(1,1);
Generator(end,end)=-nudiag(end,end);

ProbMatrix=expm(t.*Generator);%Calculating probability matrix
figure %Plotting propablity matrix
colormap gray;
cm=colormap;
colormap(flipud(cm)); %inverting colormap
ax=axes;
Pmplot=surf(ax,k,k,ProbMatrix), view(ax,2);
title(ax,['probability distribution after ',num2str(t,'%g'),' s']);
xlabel(ax,'final charge');
ylabel(ax,'start charge');
set(Pmplot,'LineStyle','none');
end
```

**Example Figure 1:**

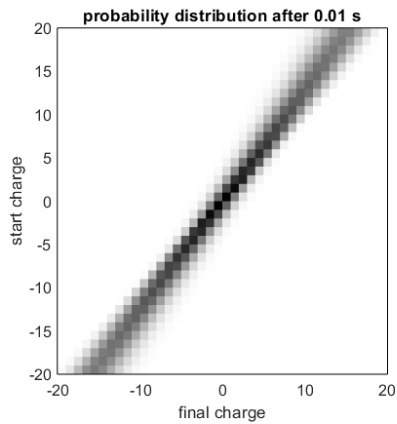

2. Matlab function to calculate the time evolution of the bipolar charging process for a given Generator, arrays of Initial charge distribution (IniCD), charge number range (k) and time range (t):

```matlab
function [FinalCD,Ergodiclimit]=TimeEvolutionCalculator(Generator,IniCD,k,t)
%% calculate the ergodic limit in two ways
%% Way One: The ergodic limit is given by the kernel of the transposed Generator
%% Way Two: THe ergodic limit is given by the eigenvalue of the transposed generator that
%% corresponds to the Eigenvalue 0.

kernel=null(transpose(Generator));
Ergodiclimit=kernel/sum(kernel);   %Normalisation
```

```matlab
[V,D,W] = eig(transpose(Generator));
Eigenvalues=D(1:(length(D)+1):length(D)^2);
lifetimes=-1*(1./Eigenvalues);%Extracting apparent lifetimes of the E-functions
[~,index]=min(abs(Eigenvalues));%Finding the Eigenvalue closest to zero
Eigenvector2Null=V(:,index)/sum(V(:,index));%Normalisation

%% Calculate Final Charge Distribution at various charging times
for i=1:length(t)
        FinalCD(:,i)=sum(diag(IniCD)*expm(t(i)*Generator),1);
end

figure('position',[100 100 1200 600])
colormap gray;
cm=colormap; colormap(flipud(cm))

%%Plot Initial Charge distribution
ax1=axes('Position',[0.08,0.08,0.2,0.8]);
plot(k,100*IniCD,'k')
set(ax1,'YLim',[-5,100])
ylabel('Particle Charge Fraction / %');
xlabel('start charge');
title('Start Charge Distribution(t=0)')
legend({'Initial Charge Distribution'},'Location','north')

%% Plot Time Evolution
ax2=axes('Position',[0.36,0.08,0.28,0.8]);
set(ax2,'CLim',[0,1]);
timeev=surf(t,k,FinalCD*100);view(2);
set(timeev,'LineStyle','none');
set(ax2,'XScale','log','Box','on','CLim',[0,100])
c=colorbar;
title(c,'% Charge')
xlabel('t / s');
ylabel('Charge');
title('Charge Distribution Time Evolution')
r=annotation(gcf,'rectangle',get(ax2,'Position'))

%% Plot Limit Charge Distributions
ax3=axes('Position',[0.72,0.08,0.2,0.8]);
hold on
        plot(k,100*FinalCD(:,end),'k')%Last Time Point
        plot(k,100*Ergodiclimit,'--r')%Ergodic Limit Way 1: Kernel
        plot(k,100*Eigenvector2Null,'xb')%Ergodic Limit Way 2: Eigenvector
hold off
box on
ylabel('Particle Charge Fraction / %');
set(ax3,'YLim',[-5 50])
xlabel('Final Load');
legend({'Final Charge Distribution','Ergodic Limit','Eigenvector (Eigenvalue=Null)'},…
…'Location','north')
title('Final Charge Distribution (t=\infty)')
end
```

**Example figure 2:**

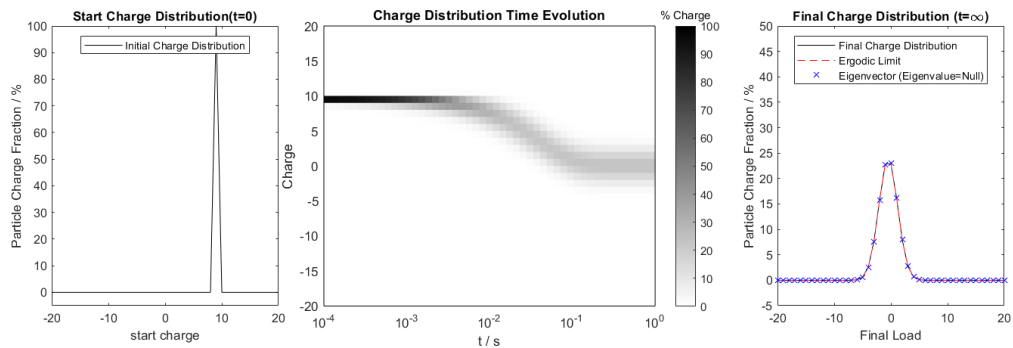

3. **Matlab function to generate analytic ion uptake coefficients for particles of a given diameter (d), charge (k) and ion mobility for positive (mobipos) and negative (mobineg) ions. For more realistic calculations, the array of uptake coefficients can be replaced by experimentally validated uptake coefficients for positive (mu) and negative ions (nu).**

```matlab
function markov_code_example_starter

%Parameters
n=41; %total dimension of state space (e.g. -20 to +20 elementary charges)
k=-((n-1)/2):((n-1)/2); %array of possible aerosol charges (e.g. -20 to +20 charges)
e=1.602176634e-19;%natural constant: elementary charge
epsilon0=8.8541878128e-12; %natural constant: dielectric permittivity
kb=1.380649e-23;%natural constant: boltzmann constant
d=300e-9;   %particle diameter =300 nm
mobipos=1.35e-4;%electrical mobility of positive ions taken from Wiedensohler 1986
mobineg=1.6e-4;%electrical mobility of negative ions taken from Wiedensohler 1986
n_neg=1e13; %high concentration of negative ions in bipolar neutralisator
n_pos=1e13; %high concentration of positive ions in bipolar neutralisator

%uptake coefficient depeding on aerosol charge (Gunn '54)

lambda=e^2/(4*pi*epsilon0*d*kb*300); %parameter Lambda
mu=k.*e*mobipos./(epsilon0*(exp(2*lambda.*k)-1));
mu((n+1)/2)=e*mobipos/(2*epsilon0*lambda);
nu=k.*e*mobineg./(epsilon0*(1-exp(-2*lambda.*k)));
nu((n+1)/2)=e*mobineg/(2*epsilon0*lambda);

%Anal. distr. Clement & Harrison 91 (only for comparison)
AnalyticCD=(n_pos*mobipos/n_neg/mobineg).^k.*sinh(lambda*k).*exp(-lambda*k.^2)./(lambda*k);
AnalyticCD((n+1)/2)=1;AnalyticCD=AnalyticCD/sum(AnalyticCD);% Normalization

t=0.01;%charging duration
%% Calcuate generator
[ProbMatrix,Generator]=GeneratorConstructionFunction(n_pos,n_neg,mu,nu,k,t);

Time evolution of an initial distinct charge distribution
IniCD=zeros(size(k));IniCD(30)=1; %initial charge distribution
t=logspace(6,10);                  %Time Point Vector
[FinalCD,Ergodiclimit]=TimeEvolutionCalculator(Generator,IniCD,k,t)
```



\end{document}